\documentclass[12pt]{article}
\usepackage{amssymb,amsmath,amsfonts}
\usepackage{hyperref}

\newcommand{\bbbone}{\mathchoice {\rm 1\mskip-4mu l} {\rm 1\mskip-4mu l}{\rm 1\mskip-4.5mu l} {\rm 1\mskip-5mu l}}

\newcounter{resultcounter}[section]

\newtheorem{thm}[resultcounter]{Theorem}
\newtheorem{lem}[resultcounter]{Lemma}
\newtheorem{prop}[resultcounter]{Proposition}

\newtheorem{definition}[resultcounter]{Definition}

\def\bed{\begin{definition}}
	\def\eed{\end{definition}}

\begin{document}
	\title{Qubit dynamics with classical noise}

\author{ Qin Huang\footnote{qinh@mun.ca} \qquad Marco Merkli\footnote{merkli@mun.ca, \href{https://www.math.mun.ca/merkli/}{website}}\\
	\ \\
Department of \\
 Mathematics and Statistics\\
 Memorial University of Newfoundland\\
 St. John's, NL\\
 A1C 5S7\\
  Canada}

	\maketitle

\begin{abstract}
We study the evolution of a qubit evolving according to the Schr\"odinger equation with a Hamiltonian containing  noise terms, modeled by random diagonal and off-diagonal matrix elements. We show that the noise-averaged qubit density matrix converges to a final state, in the limit of large times $t$. The convergence speed is polynomial in $1/t$, with a power depending on the regularity of the noise probability density and its low frequency behaviour. We evaluate the final state explicitly. We show that in the regimes of weak and strong off-diagonal noise, the process implements the dephasing channel in the energy- (localized) and the delocalized basis, respectively.
\end{abstract}
	
\section{Introduction}

The state of a quantum system having two degrees of freedom, that is, a spin or a qubit, is given by a {\em density operator} $\rho$ acting on ${\mathbb C}^2$, which is positive definite and has trace one, and is represented by a $2\times 2$ matrix. States evolve according to the Schr\"odinger, or Liouville - von Neumann equation,
\begin{equation}
		\label{schro}
		i\partial_t\rho(t)=[H_0,\rho(t)],
\end{equation}
where $H_0$ is the Hamiltonian, a fixed hermitian operator acting on ${\mathbb C}^2$.  The subindex $0$ indicates that this is a `noiseless' Hamiltonian. Given any initial state $\rho(0)$, the equation \eqref{schro} has the unique solution 
	\begin{equation}
		\label{schro1}
		\rho(t)=e^{-itH_0}\rho(0)e^{itH_0}.
	\end{equation}
	Upon diagonalizing the Hamiltonian, we have
	\begin{equation}
		\label{H_0}
		H_0=\begin{pmatrix}
			E_1 & 0\\
			0 & E_2
		\end{pmatrix},
	\end{equation}
	with real eigenvalues $E_1$, $E_2$ and associated eigenvectors 
	\begin{equation}
		\label{Phi}
		\Phi_1 = \begin{pmatrix} 1 \\ 0\end{pmatrix},\quad 
		\Phi_2 = \begin{pmatrix} 0 \\ 1\end{pmatrix},
	\end{equation} 
	making up the so-called {\em  energy, localized or diabatic basis}. Denoting the spectral projections by $P_j=| \Phi_j\rangle\langle\Phi_j|$, $j=1,2$,  the evolution \eqref{schro1} takes the form
	\begin{equation}
		\rho(t)= \sum_{k,l\in\{1,2\}} e^{-i t(E_k-E_l)} P_k\rho(0)P_l\ .
		\label{m4'}
	\end{equation}
The density matrix evolves entirely {\em coherently}, namely as a superposition of time-periodic functions. This is the fate of any (finite-dimensional) quantum system evolving according to the Schr\"odinger equation. But this is {\em not} what is often observed in nature:  Commonly systems undergo irreversible processes, such as equilibration and decoherence as well as transport processes, like the exchange of excitation and charges within molecules. These phenomena are {\em noise effects} induced by the contact with external agents.
	
As an example, {\em decoherence}, equivalently called {\em dephasing}, is an important phenomenon in  quantum theory and in the quantum information sciences in particular.  To describe it, we denote the density matrix elements in the energy basis by
	\begin{equation}
		\rho_{ij}(t)=\big\langle \Phi_i,\rho(t)\Phi_j\big\rangle.
		\label{m4}
	\end{equation}
Many quantum effects, which can be exploited in quantum computing for example,  come from the fact that a pure state can be a {\em superposition} of the energy eigenvectors $\Phi_j$, {\em i.e.}, $\rho = |\Psi\rangle\langle \Psi|$ with $\Psi= a_1\Phi_1 +a_2\Phi_2$ and $|a_1|^2+|a_2|^2=1$. 
More generally, any mixed state is of the form
	\begin{equation}
		\label{mixed}
		\rho = p |\Phi_1\rangle\langle\Phi_1|+ (1-p) |\Phi_2\rangle\langle\Phi_2|+ z |\Phi_1\rangle\langle\Phi_2| + \bar z|\Phi_2\rangle\langle\Phi_1|,
	\end{equation}
for some $p\in[0,1]$, and where $z\in\mathbb C$, $|z|^2\le p(1-p)$, is called the coherence of $\rho$. (In the pure state case  $z=a_1\bar a_2$.) When $z=0$, we call the state $\rho$, \eqref{mixed}, an {\em  incoherent superposition} of the energy states. For $z\neq 0$ $\rho$, \eqref{mixed}, is called a {\em coherent superposition} of $\Phi_1$ and $\Phi_2$.\footnote{This notion should not be confused with that of a ``coherent state'', say, of an oscillator or similar.} The average  measurement value  when measuring an observable ${\mathcal O}$ in the state $\rho$, \eqref{mixed}, is given by
	\begin{equation}
		\label{average}
		\langle {\mathcal O}\rangle = {\rm tr} \rho {\mathcal O}= p \langle {\mathcal O}\rangle_1 + (1-p) \langle {\mathcal O}\rangle_2  + 2{\rm Re }  \,z \langle \Phi_2, {\mathcal O}\Phi_1\rangle,
	\end{equation}
where $\langle{\mathcal O}\rangle_j=\langle\Phi_j, {\mathcal O}\Phi_j\rangle$, $j=1,2$. For an incoherent $\rho$, the last term on the right side of \eqref{average} vanishes (as $z=0$) and the measurement process has the characteristics of a classical  system, where two states with given energy are mixed with probabilities $p$ and $1-p$. However, for a coherent $\rho$, the cross term $2{\rm Re }  \,z \langle \Phi_2, {\mathcal O}\Phi_1\rangle$ gives an additional term, which reflects an interplay of $\Phi_1$ and $\Phi_2$ within the state $\rho$. This term is due to the quantum nature of $\rho$ -- two energy states are mixed in a quantum way to give a state $\rho$. Note that for observables $\mathcal O$ commuting with $H_0$, this effect is not visible, as $\langle \Phi_2,{\mathcal O}\Phi_1\rangle=0$.  Coherence is a basis dependent notion because of course, every density matrix can be diagonalized, $\rho=p|\chi_1\rangle\langle\chi_1| +(1-p)|\chi_2\rangle\langle\chi_2|$, for eigenvectors $\chi_{1,2}$ and some $p\in[0,1]$. So $\rho$ is always an incoherent superposition of its eigenstates. However, those eigenstates may not have a direct physical meaning in the sense that they might not be eigenvectors of an easily accessible observable. 
	
The process of {\em decoherence} (in the energy basis) is defined to be the transition of $\rho\mapsto \rho'$ were $\rho'$ is an incoherent superposition of the energy states. In other words, it is a process in which the off-diagonal density matrix elements of $\rho$, written in the energy basis, disappear. In quantum information theory, a {\em channel} $\mathcal E$ is defined as a completely positive, trace preserving map acting on density matrices. The {\em dephasing channel} is given by \cite{Preskill} 
	\begin{equation}
		\label{dechannel}
		{\mathcal E}\rho = {\mathcal E}
		\begin{pmatrix}
			\rho_{11} & \rho_{12}\\
			\rho_{21} & \rho_{22}
		\end{pmatrix}
		=
		\begin{pmatrix}
			\rho_{11} & (1-\eta)\rho_{12}\\
			(1-\eta)\rho_{21} & \rho_{22}
		\end{pmatrix},
	\end{equation}
where $0\le \eta\le 1$ controls the reduction of the off-diagonals in the energy basis. For $\eta=1$ the coherences are entirely suppressed by the action of $\mathcal E$ and the state undergoes full decoherence.  It is well known (and discussed in a huge amount of literature) that generically, quantum systems undergo a gradual reduction of coherences when they are in contact with environments \cite{Zurek, Joos, BP, PSE}. In this sense, decoherence is a generically observed dynamical process, brought about by external noises. Using \eqref{m4}, we express (full) decoherence as
	\begin{equation}
		\label{decoprocess}
		\rho_{12}(t) \rightarrow 0 \quad \mbox{as $t\rightarrow\infty$.}\footnote{If the limit of the off-diagonals converge to zero we say there is {\em full decoherence}. Depending on the models, one also observes {\em partial decoherence}, where the limit value of the off-diagonal is reduced relative to the initial value, but possibly nonzero \cite{PSE, MBSWN}.}
	\end{equation}
Decoherence plays a central role in quantum theory, particularly in quantum information and computation, where coherence is a resource exploited in the design of fast algorithms, and decoherence is to be avoided as much as possible. A core task is to establish mathematical models which enable to uncover mechanisms driving the decoherence process, and quantify them by, for instance, finding the speed of convergence in \eqref{decoprocess}. This is a first step in designing countermeasures to protect systems from losing quantum features due to noise effects. 
	\smallskip

It results from the above discussion that it is an important task  to find the dynamical equation of a quantum system showing irreversible characteristics. This has to be done by including noise effects, {\em i.e.}, by considering {\em open quantum systems}. Their dynamics is {\em not} given `simply' by the Schr\"odinger equation \eqref{schro}. The latter only describes the dynamics of {\em closed} systems. In the literature, one can find the following approaches to describing open system dynamics.

\begin{itemize}
\item[(H)] {\bf Hamiltonian approach.} The system of interest (say a qubit) is coupled to an `environment', a very large  other quantum system having many degrees of freedom. The complex (system + environment) is regarded as a closed system and evolves according to the Schr\"o\-dinger equation, governed by a Hamiltonian which describes the system, the environment and the interaction between the two. In a sense, this is the most fundamental description of a noisy system, but at the same time it is enormously complicated, because the dynamics describes all the details about the system and all degrees of freedom of the reservoir \cite{BP,MSB, GZ, Davies, KoMe, KoMeSo, JP}. 
		
Upon restricting the full dynamics to just the system, by `tracing out the environment degrees of freedom', one arrives at an effective equation for the system alone. In the absence of interaction with the environment, this equation reduces to the system Schr\"odinger equation, but it is much more complicated in the presence of interactions. In certain approximative regimes (weak coupling, fast reservoir dynamics or dissipation), this effective equation takes the form of the ubiquitous {\em Markovian master equation}. A rigorous derivation of the master equation has recently been given in \cite{MAOP}.

\item[(E)] {\bf Effective evolution.} The Markovian master equation resulting from the procedure explained in (H) is of the form $\rho(t)=e^{t{\mathcal L}}\rho(0)$. Here, the Lindblad operator $\mathcal L$ (which acts on density matrices) is not hermitian and has complex eigenvalues $-x+iy$ (with $x\ge 0$), leading to time decay $\sim e^{-tx}$, thus describing irreversible effects. Moreover, since $e^{t{\mathcal L}}$ is a group of completely positive, trace preserving maps, its generator $\mathcal L$ is constrained to have  a standard form (Lindblad, Gorini-Kossakovski-Sudarshan) \cite{CP, AL, DS}. This standard form is often taken as the {\em starting point} for modeling an open system dynamics. Namely, one specifies the components in the standard form to build a generator $\mathcal L$ without deriving it form a microscopic model as in (H). Then one analyzes the Markovian dynamics resulting from $\mathcal L$ directly  \cite{Haroche}.

\item[(R)] {\bf Random Hamiltonians.} Another way of describing noise effects is to incorporate  randomness in the Hamiltonian of the system. One can adopt two points of view: 
		
\quad (i) The system is in contact with a time-dependent noise source, turning the Hamiltonian into a stochastic process, or 
		
\quad (ii) The system Hamiltonian is a  time-independent random matrix. 
		
In (i) one imagines that the environment acts as an external agent, randomly shaking the system at all times \cite{Nesterov,Aguilar}. In (ii), the effect of the reservoir is to randomize the energies and interaction parameters of the system. Each realization of the Hamiltonian leads to a deterministic dynamics and a statistics is built up by observing the evolution over many realizations. This is in line with the point of view taken in models governed by random Schr\"odinger equations, as for instance the Anderson model. {\em In the present work, we adopt this view.} Taking the expectation of the density matrix is the equivalent of tracing out the reservoir degrees of freedom in (H) above and will yield time-decay and irreversible dynamical effects. In this context some authors have considered systems coupled to mesoscopic environments, modeled by random matrix Hamiltonians of many levels \cite{BratusPastur}. In contrast, we incorporate the environment noise as random energy terms directly into the  Hamiltonian.

	\end{itemize}

	\section{The Model}
	
	The Hamiltonian of the system is 
	\begin{equation}
		H = H_0+H_{\rm noise},
		\label{H}
	\end{equation}
	where
	\begin{equation} 
		H_0= 
		\begin{pmatrix}
			E_1 & 0 \\
			0 & E_2 \\
		\end{pmatrix}
		\quad\mbox{and}\quad 
		H_{\rm noise}=
		\begin{pmatrix}
			\xi_1 & \xi_{\rm o} \\
			\xi_{\rm o} & \xi_2 \\
		\end{pmatrix}.
		\label{hamiltonian2}
	\end{equation}
	Here, $E_1, E_2 \in\mathbb R$ are constants and $\xi_{\rm o}$, $\xi_1$ and $\xi_2$ are real valued, independent random variables representing the noise. We assume that the Bohr energy satisfies
	\begin{equation}
		\label{eps}
		\varepsilon = E_1-E_2 >0
	\end{equation}
	and define the quantity
	\begin{equation}
		\label{xid}
		\xi_{\rm d} = \xi_1-\xi_2.
	\end{equation}
We call $\xi_{\rm d}$ and $\xi_{\rm o}$ the diagonal and the off-diagonal (or, tunneling) noise, respectively. To those random variables are associated the probability densities\footnote{The probability of the event $\{\xi\in[a,b]\}$ is given by $\int_a^b \mu(x)dx$.}
\begin{equation}
	\label{probadensity}
\xi_{\rm d} \leftrightarrow \mu_{\rm d}(y)\qquad \mbox{and}\qquad \xi_{\rm o} \leftrightarrow \mu_{\rm o}(x).
\end{equation}

The eigenbasis of $H_0$ is denoted by \eqref{Phi}.
Shifting the Hamiltonian $H$ by adding a matrix $\alpha \bbbone$, where $\alpha$ is a real number (or a random variable) and $\bbbone$ is the $2\times 2$ identity matrix, does not alter the evolution of quantum states, as $e^{-i t (H+\alpha \bbbone)}\rho\, e^{i t (H+\alpha \bbbone)} = e^{-i t H}\rho\, e^{i t H}$. It is then apparent from \eqref{hamiltonian2} that only the quantities \eqref{eps} and \eqref{xid}  will play a role in the dynamics.

Open two-level (two state) systems are ubiquitous in quantum theory. Despite being of mathematically simplest form (two-dimensional!), they represent diverse physical systems, ranging from qubits, spins or atoms interacting with radiation in quantum information theory and quantum optics \cite{Haroche,Kreinberg} to  donor-acceptor systems in quantum chemical and quantum biological processes \cite{XuSchulten,MDimer}. The analysis of open two-level systems is far from trivial \cite{Leggett} and new results are emerging regularly \cite{Ferialdi, KoMe,KoMeSo}. One possible  realization of such a two-level system is given by a quantum particle in a double well potential, having minima (say at spatial locations $x_1$ and $x_2$). It is assumed that the wells are deep enough so that it makes sense to talk about the states $\Phi_1$ and $\Phi_2$ representing the particle being located in the respective well. The associated energies are $E_1$ and $E_2$ and they are shifted by $\xi_1$ and $\xi_2$ due to external noise. The particle can tunnel between the wells due to environmental effects, which corresponds to the tunneling matrix element $\xi_{\rm o}$ in \eqref{hamiltonian2}. 
In contrast to the localized basis \eqref{Phi}, one introduces the `delocalized' or `adiabatic' basis \cite{May} given by
\begin{equation}
		\label{delocbasis}
		\Phi_\pm =\frac{1}{\sqrt 2} (\Phi_1\pm \Phi_2).
\end{equation}
If $\Phi_{1,2}$ represent states with relatively well localized positions,  then $\Phi_\pm$ are those having the largest position uncertainty (variance), so they are the `most delocalized'.

\subsection{Highlights of our main results}

The goal of this section is to outline our main results without much technicality. We present the mathematical details in Section \ref{mainresultsect} below.

We assume that the probability densities $\mu_{\rm d}$ and $\mu_{\rm o}$, \eqref{probadensity} are compactly supported within intervals $(-\eta_{\rm d}, \eta_{\rm d})$ and $(-\eta_{\rm o}, \eta_{\rm o})$, respectively. We also assume that $\eta_{\rm d} <\varepsilon$. The density matrix $\rho(t)$ is a random matrix (because the Hamiltonian $H$ is) and we denote the average with respect to the randomness by ${\mathbb E}[\rho(t)]$ (`expectation'). More precisely, throughout the paper, if $F\equiv F(x,y)$ is a random quantity depending on the off-diagonal noise variable $x$ and the diagonal one $y$ (see also \eqref{probadensity}), we denote its average as
\begin{equation}
	\label{notation}
	{\mathbb E}[F]=\int_{{\mathbb R}^2} F(x,y) \mu_{\rm o}(x) \mu_{\rm d}(y) dx dy . 
\end{equation}
Our main findings are these.
\begin{itemize}
\item[{\bf 1.}]
{\bf Diagonal noise drives system to final state.}  We show in Theorem \ref{theorem1} that if $\mu_{\rm d}$ is $n=1,2,\ldots$ times continuously differentiable, then for $t\ge 0$,
\begin{equation}
	\label{r1}
\big\| {\mathbb E}[\rho(t)] - \bar\rho \ \big\| \le \frac{C}{1+t^n}.
\end{equation}
Here, $\bar\rho$ is an explicit final state which depends on the noises and on the initial state $\rho(0)$. The result also holds in the absence of the off-diagonal noise $\xi_{\rm o}$. This shows  that the diagonal noise drives the state to a final one. {\em The speed of convergence depends on the smoothness of the noise distribution: the higher the smoothness, the quicker the convergence.}

\item[{\bf 2.}] {\bf Off-diagonal noise: Low energy modes slow down approach of final state.} In the absence of diagonal noise, when only $\xi_{\rm o}$ is present, our Theorem \ref{theorem1'} shows that the diagonals ${\mathbb E}[\rho_{jj}(t)]$ converge to $\bar\rho_{jj}$ at speed $1/t$. Moreover, if the initial coherence vanishes, $\rho_{12}(0)=0$, then \eqref{r1} holds with $n=1$. {\em Increased regularity of $\mu_{\rm o}$ does not speed up the convergence, however.} The hindrance to a speedup are the slow noise modes (frequencies close to $0$): We show that when the latter are suppressed, meaning that $\mu_{\rm o}(\omega)$ vanishes at the origin as $\mu_{\rm o}(\omega)\sim \omega^k$ for some $k=1,3,5,\ldots$  then the convergence \eqref{r1} is valid with $n=\frac{k+1}{2}$.

\item[{\bf 3.}] {\bf Dynamics implements dephasing channel.} We establish the explicit form of the final state $\bar\rho$ in all parameter regimes. The relation (initial state) $\rightarrow$ (final state) is particularly easy and elegant in two regimes:
\begin{itemize} 
	\item[(i)] For weak off-diagonal noise, characterized by $\eta_{\rm o} <\!\!< \varepsilon$, 
	\item[(ii)] For strong off-diagonal noise,  $\varepsilon <\!\!<\mu_{\rm o}^{\rm min}\equiv \min \{ |\omega|\, : \, \mu_{\rm o}(\omega) \neq 0\}$.
\end{itemize}
In the case (i), the maximal frequency of the noise is small compared to the Bohr frequency $\varepsilon$ of the qubit and in the case (ii) the noise has only frequencies much larger than $\varepsilon$. We show that in these two regimes, the final state is given by
\begin{equation*}
	\bar\rho =  \left\{
	\begin{array}{cl}
		\rho_{11}(0)\, |\Phi_1\rangle\langle \Phi_1| + \rho_{22}(0)\,  |\Phi_2\rangle\langle\Phi_2|  +O(\eta_{\rm o}/\varepsilon)& \quad \mbox{weak off-diag. noise}\bigskip\\
		\rho_{++}(0)\,  |\Phi_+\rangle\langle \Phi_+| +\rho_{--}(0)\,  |\Phi_-\rangle\langle \Phi_-| +O(\varepsilon/\mu_{\rm o}^{\rm min})& \quad \mbox{strong off-diag. noise}
	\end{array}
	\right.
\end{equation*}
where $\Phi_{1,2}$ is the localized basis \eqref{Phi} and $\Phi_\pm$ is the delocalized basis \eqref{delocbasis}. We also set $\rho_{++}=\langle \Phi_+,\rho\,  \Phi_+\rangle$ and analogously for $\rho_{--}$. 

{\em This shows that the noise implements the {\em dephasing channel} \eqref{dechannel} (with $\eta=1$) in the localized basis (weak noise) or in the delocalized basis (strong noise).} 

The speed at which the channel is implemented depends on the properties of $\mu_{\rm d}$ and $\mu_{\rm o}$, as specified in the results on convergence given above in points 1. and 2.
\end{itemize}

\subsection{Heuristic analysis}

We find it instructive to present a heuristic derivation of our results before giving the rigorous analysis in Section \ref{mainresultsect}. The first step in our analysis is the explicit diagonalization of $H$, \eqref{H}. The details are done in  Section \ref{proofsect}, and the result is Lemma \ref{lemma3}. For the heuristic analysis here, we require only the following information, also contained in this lemma: The density matrix elements are of the form
\begin{equation}
\label{h1}
\rho_{kl}(t) = \bar\rho_{kl} (x/\varepsilon, y/\varepsilon)+ p_{kl}(x/\varepsilon, y/\varepsilon)\, e^{\pm i t\varepsilon \Phi(x/\varepsilon, y/\varepsilon)}
\end{equation}
with  $\xi_{\rm o}(x)=x$ and $\xi_{\rm d}(y)=y$ and where the dynamical phase is given by 
\begin{equation}
\label{h1.1}
\Phi(x/\varepsilon, y/\varepsilon) = (1+y/\varepsilon)\sqrt{1+4\frac{(x/\varepsilon)^2}{(1+y/\varepsilon)^2}} \ .
\end{equation}
Throughout, $x$ denotes the off-diagonal, $y$ the diagonal noise variable, see also \eqref{probadensity}.  In \eqref{h1.1}, $\bar\rho_{kl}$ and $p_{kl}$ are both quantities depending on the initial state $\rho(0)$ and the noises, but they are time independent.  The $\pm$ in \eqref{h1} indicates that a linear combination can occur (with different $p_{kl}$). The average over $x$ and $y$ of the exponential  carrying the dynamical phase in \eqref{h1} determines the time decay properties of $\rho_{kl}(t)$. 

\subsubsection{Heuristics of the weak noise regime}  This is the regime in which the probability densities $\mu_{\rm o}(x)$ and $\mu_{\rm d}(y)$ yield the restrictions $x/\varepsilon$, $y/\varepsilon <\!\!<1$. Up to order three in the noise, we have from \eqref{h1.1} 
\begin{equation}
\label{h2}
\Phi(x/\varepsilon, y/\varepsilon) \ \sim\   1 +  y/\varepsilon +2 (x/\varepsilon)^2.
\end{equation}
Depending on the values of $k,l$, the $p_{kl}$ have the lowest order expansions (Lemma \ref{analemma})
\begin{equation}
	\label{h3}
p_{kl}(x/\varepsilon, y/\varepsilon) \sim  (x/\varepsilon)^n,\qquad n=0,1,2.
\end{equation}
This implies 
\begin{equation}
\label{h4}
{\mathbb E} \Big[p_{kl}(x/\varepsilon, y/\varepsilon)e^{-i t\varepsilon\Phi(x/\varepsilon, y/\varepsilon)}\Big] \ \sim\   e^{-it\varepsilon}\,  {\mathbb E}\big[e^{-i t\varepsilon (y/\varepsilon)}\big] \, {\mathbb E}\big[(x/\varepsilon)^n e^{-2i t\varepsilon (x/\varepsilon)^2}\big].
\end{equation}
The contribution coming from the diagonal noise is given by the Fourier transform of the probability density,
\begin{equation}
\label{h5}
{\mathbb E}\big[e^{-i t\varepsilon (y/\varepsilon)}\big] =\int_{\mathbb R} e^{-i t y}\mu_{\rm d}(y) dy = \widehat \mu_{\rm d}(t). 
\end{equation}
The decay for large values of $t$ is determined by the smoothness of $\mu_{\rm d}(y)$. If $\mu_d$ is $k$ times continuously differentiable then for large $t$ we have $\widehat \mu_{\rm d}(t) \sim t^{-k}$; for a Gaussian $\mu_{\rm d}(x)\propto e^{-x^2/2\sigma^2}$, the decay is $\widehat \mu_{\rm d}(t)\propto e^{-\sigma^2t^2/2}$. 

The contribution to \eqref{h4} coming from the off-diagonal noise is
\begin{eqnarray}
\label{h6}
{\mathbb E}\big[(x/\varepsilon)^n e^{-2i t\varepsilon (x/\varepsilon)^2}\big] &=& \varepsilon \int_{\mathbb R} x^n e^{-2 i  (\sqrt{\varepsilon t} x)^2} \mu_{\rm o}(\varepsilon x) dx\nonumber\\
&=& \varepsilon t^{-(n+1)/2} \int_{\mathbb R} x^n e^{-2 i  x^2} \mu_{\rm o}\big( x\sqrt{ \varepsilon/ t}\, \big) dx\nonumber\\
&\sim& t^{-(n+1)/2}\mu_{\rm o}(0), \quad t>\!\!>\varepsilon.
\end{eqnarray}
This contribution decays as an inverse power of $\sqrt{t}$, a power which does {\em not} depend on the shape (smoothness) of $\mu_{\rm o}$, but only on the value $\mu_{\rm o}(0)$. The slowest decay is for terms with $n=0$, and is given by $1/\sqrt t$. (Even though quicker decay can be achieved by suppressing slow noise modes, {\em i.e.}, if $\mu_{\rm o}(0)=0$.) This heuristic analysis shows that:
\begin{itemize}
\item[$\bullet$] In the weak noise regime, both the diagonal and the off-diagonal noises contribute to the convergence of the density matrix to a final state $\bar\rho$,
\begin{equation}
	\label{heu1}
\big\|\,  {\mathbb E}[\rho(t)] - \bar\rho \, \big\| \ \sim\  \big| \widehat \mu_{\rm d}(t)\, \mu_{\rm o}(0)\big| \, t^{-1/2}.
\end{equation}
The convergence speed caused by the diagonal noise is given by the factor $|\widehat \mu_{\rm d}(t)|$ in \eqref{heu1}, the Fourier transform of the probability density $\mu_{\rm d}$. It therefore depends on the smoothness of the diagonal noise. In contrast, the convergence speed due to the off-diagonal noise, $\mu_{\rm o}(0) \, t^{-1/2}$, is insensitive to the shape and smoothness of $\mu_{\rm o}$.
\end{itemize}

\subsubsection{Heuristics of the strong off-diagonal noise regime}

This is the regime in which $\mu_{\rm d}(y)$ and $\mu_{\rm o}(x)$, \eqref{probadensity}, are such that $|x|/\varepsilon >\!\!>1+y/\varepsilon$. (Note that $1+y/\varepsilon>0$ due to \eqref{eps} and $\eta_{\rm d}<\varepsilon$.) The dynamical phase \eqref{h1.1} is
\begin{equation}
\label{h7}
\Phi(x/\varepsilon, y/\varepsilon) \sim  |x|/\varepsilon.
\end{equation}
According to Lemma \ref{lemma3}, in this regime we have 
\begin{equation}
\label{h8}
p_{kl}(x/\varepsilon, y/\varepsilon) \sim \mbox{constant}.
\end{equation}
Let us discuss the situation when $\mu_{\rm o}$ is supported essentially around some average value $x_*>\!\!>\varepsilon>0$ (the general case is treated in the same way). Then $|x|/\varepsilon \sim x/\varepsilon$ and 
\begin{equation}
\label{h9}
{\mathbb E} \Big[p_{kl}(x/\varepsilon, y/\varepsilon)e^{-i t\varepsilon\Phi(x/\varepsilon, y/\varepsilon)}\Big] \ \sim\   e^{-it\varepsilon}\,  {\mathbb E}\big[e^{-2i t x}\big]  = e^{-it\varepsilon}\,  \widehat \mu_{\rm o}(2t).
\end{equation}
In contrast to the weak coupling regime, here the decay depends on the smoothness of the off-diagonal noise $\mu_{\rm o}$, while the diagonal noise does not contribute to the convergence at all. This heuristic analysis shows:
\begin{itemize}
\item[$\bullet$] 
In the strong off-diagonal noise regime, only the off-diagonal noise contributes to the decay of the density matrix towards a final state $\bar \rho$,
	$$
\big\|\, {\mathbb E}[\rho(t)] - \bar\rho \, \big\| \ \sim\   \big| \widehat \mu_{\rm o}(2t)\big|.
	$$ 
The time-decay of the Fourier transform $\widehat\mu_{\rm o}(2t)$ depends on the smoothness of $\mu_{\rm o}$, the probability density of the off-diagonal noise. 

\end{itemize}

A heuristic identification the final state $\bar\rho$ is seems more difficult. In particular, the final state depends on the initial condition $\rho(0)$. Nevertheless, for this simple $2 \times 2$ system, it can be calculated explicitly, see Theorems \ref{theorem1} and \ref{theoremfinalstate}.

\section{Main results, rigorous}
\label{mainresultsect}

In order to make the analysis rigorous, we make the following assumption.
\medskip

{\bf Assumption (A)} The probability densities $\mu_{\rm o}$ and $\mu_{\rm d}$, \eqref{probadensity}, have compact support in the open intervals $(-\eta_{\rm o}, \eta_{\rm o})$ and $(-\eta_{\rm d}, \eta_{\rm d})$, respectively, where  $0<\eta_{\rm o}<\infty$ and $0<\eta_{\rm d}<\varepsilon$. Recall the notation \eqref{notation} for the expectation $\mathbb E$.

\begin{thm}[Convergence]
\label{theorem1}
Suppose assumption {\em (A)} holds and that $\mu_{\rm d}$ is $n$ times continuously differentiable for some $n\in{\mathbb N}\cup\{0\}$. Then there is a constant $C_n$  {\em s.t.}  for all $t\ge0$,
\begin{equation}
\big\| \, {\mathbb E}[\rho(t)]- \bar \rho\,  \big\| \le \frac{C_n}{1+t^n}.
	\label{mainres}
	\end{equation}
The final state $\bar \rho$ is given by
\begin{eqnarray}
\label{finalstate}
\bar\rho_{11}  &=&  \alpha +\beta\rho_{11}(0)-2\gamma {\rm Re} \rho_{12}(0)\nonumber\\
\bar\rho_{12} &=& \gamma (1-2\rho_{11}(0)) + 2\alpha {\rm Re}\rho_{12}(0),
\end{eqnarray}
where $\alpha,\beta\ge 0$, $\gamma\in\mathbb R$, are explicit constants depending on the noises but not on the initial state $\rho(0)$. Moreover, if the off-diagonal noise satisfies $\mu_{\rm o}(-x) = \mu_{\rm o}(x)$, then $\gamma =0$.
\end{thm}
The explicit formulas for $\alpha,\beta,\gamma$ are given in \eqref{52}.  The bound \eqref{mainres} is consistent with the heuristic estimate \eqref{heu1}. We note though, that decay of non-integer powers (such as $t^{-1/2}$ as in \eqref{heu1}) is not detected in Theorem \ref{theorem1}. This is because we derive the result using integration by parts, which only yields decay of integer inverse powers of $t$.

Theorem \ref{theorem1} is also valid in case the off-diagonal noise vanishes, {\em i.e.}, for $\xi_{\rm o}=0$ in \eqref{hamiltonian2}. We conclude that the diagonal noise drives the convergence to a final state $\rho$, at a speed depending on the smoothness of the noise distribution. Our next result examines the situation when $\xi_{\rm d}=0$. 

\begin{thm}[Purely off-diagonal noise]
	\label{theorem1'}
	Suppose that $\xi_{\rm d}=0$ and that $\mu_{\rm o}$ is compactly supported and continuously differentiable. 
	\begin{itemize}
		\item[{\rm 1.}] There is a constant $C$ such that for all $t\ge 0$ and  $j=1,2$, 
		$$
		\big| \,{\mathbb E}[\rho_{jj}(t)] -\bar\rho_{jj} \, \big| \le \frac{C}{1+t},
		$$ 
		where $\bar \rho_{jj}$ are the diagonal elements of $\bar\rho$, \eqref{finalstate}. Moreover, if the initial density matrix is incoherent, $\rho_{12}(0)=0$, then there is a constant $C$ such that for all $t\ge 0$,
		$$
		\big\| \,{\mathbb E}[\rho(t)] -\bar\rho \, \big\| \le \frac{C}{1+t},
		$$ 
		where $\bar\rho$ is given in \eqref{finalstate}.
		
		\item[{\rm 2.}] Let $k=1,3,5\ldots$ be a fixed odd number and assume that $\mu_{\rm o}$ is $k$ times continuously differentiable and has a zero of order at least $k$ at the origin, meaning that 
		$$
		\lim_{\omega\rightarrow 0}\frac{\mu_{\rm o}(\omega)}{|\omega|^k}<\infty.
		$$
		Then there is a constant $C$ such that for all $t\ge 0$,
		\begin{equation}
			\big\| \, {\mathbb E}[\rho(t)]- \bar \rho\,  \big\| \le \frac{C}{1+t^{\frac{k+1}{2}}}\, .
			\label{mainres'}
		\end{equation}
		The final state $\bar \rho$ is given by \eqref{finalstate}.
	\end{itemize}
\end{thm}

Theorem \ref{theorem1'}  shows that the dynamical process is slowed down by slow noise modes. Namely, $\mu_{\rm o}$ has to vanish quickly at $\omega=0$ to increase the speed.
\bigskip

We point out that the final state $\bar\rho$ (the same in both Theorems \ref{theorem1} and \ref{theorem1'}) is known for all parameter regimes. The following result is obtained by expanding the coefficients $\alpha, \beta, \gamma$ (see \eqref{52}) in two regimes, where the off-diagonal noise is either small or large. Recall the meaning of $\mathbb E$ given in \eqref{notation}.

\begin{thm}[Final state]
	\label{theoremfinalstate} 
Under the assumptions of Theorem \ref{theorem1}, we have the following.
\begin{itemize}
\item[1.] {\em\bf Weak noise.} In the weak off-diagonal noise regime, defined by
\begin{equation}
	\label{weaknoise}
\nu_1\equiv \frac{\eta_{\rm o} }{\varepsilon}\frac{1}{1-\eta_{\rm d}/\varepsilon}  <\!\!< 1,
\end{equation}  
we have 
\begin{eqnarray}
\alpha &=& 2 {\mathbb E} \big[ (x/\varepsilon)^2 \big] \, {\mathbb E}\Big[ \frac{1}{(1+y/\varepsilon)^2}\Big] +O(\nu_1^4) \nonumber\\
\beta &=& 1 +O(\nu_1^2)\nonumber\\
\gamma &=&  - {\mathbb E} [ x/\varepsilon ] \, {\mathbb E}\Big[ \frac{1}{1+y/\varepsilon}\Big] +4{\mathbb E}\Big[\big(x/\varepsilon \big)^3\Big] \, {\mathbb E}\Big[ \frac{1}{(1+y/\varepsilon)^3}\Big]+O(\nu_1^5).\ \ \ 
\label{fs4}
\end{eqnarray}

\item[2.] {\em\bf Strong off-diagonal noise.} Suppose $\mu_{\rm o}$ is supported in $|x|>\mu_{\rm o}^{\rm min}$, for some $\mu_{\rm o}^{\rm min} >0$. In the strong off-diagonal noise regime, defined by 
\begin{equation}
\label{strongnoise}
\nu_2\equiv \frac{\varepsilon}{\mu_{\rm o}^{\rm min}} <\!\!<1,
\end{equation} 
we have 
\begin{eqnarray}
	\alpha &=& \frac12 +O\big(\nu_2^2\big) \nonumber\\
	\beta &=& \frac14  {\mathbb E} \Big[\frac{1}{( x/\varepsilon)^2}\Big] \, {\mathbb E}\Big[(1+y/\varepsilon)^2\Big] +O\big( \nu_2^3\big)\nonumber\\
\gamma&=& - {\mathbb E}\Big[\frac{1}{x/\varepsilon}\Big] \, {\mathbb E}\Big[ 1+y/\varepsilon\Big]+O\big(\nu_2^2\big).\ \ \ 
	\label{fs5}
\end{eqnarray}

\end{itemize}
\end{thm}

The result \eqref{fs5} for the strong noise regime holds also (approximately) if $\mu_{\rm o}$ is not strictly supported in $|x|>\mu_0^{\rm min}$. It suffices that most of the support of $\mu_{\rm o}$ be in that region. This modification is easy to quantify.
\newpage

\noindent
{\bf Discussion and Remarks} 
\begin{itemize}
\item[--] {\bf Implementation of the dephasing channel.} Relations \eqref{finalstate} and \eqref{fs4}, \eqref{fs5} show that 
\begin{equation}
\bar\rho = \left\{
\begin{array}{cl}
\rho_{11}(0)\, |\Phi_1\rangle\langle \Phi_1| + \rho_{22}(0)\,  |\Phi_2\rangle\langle\Phi_1| + O(\nu_1) & \quad \mbox{weak noise}\bigskip\\
\rho_{++}(0)\,  |\Phi_+\rangle\langle \Phi_+| +\rho_{--}(0)\,  |\Phi_-\rangle\langle \Phi_-| +O(\nu_2)& \quad \mbox{strong off-diagonal noise}
\end{array}
\right.
\end{equation}
where $\nu_{1,2}<\!\!<1$ are defined in \eqref{weaknoise}, \eqref{strongnoise} and $\Phi_{12}$ is the canonical (localized, diabatic) basis \eqref{Phi} and $\Phi_\pm =\frac{1}{\sqrt 2} (\Phi_1\pm \Phi_2)$ is the delocalized (adiabatic) basis. We also set $\rho_{++}=\langle \Phi_+,\rho\,  \Phi_+\rangle$ and analogously for $\rho_{--}$. This shows that the noise implements the {\em dephasing channel} in the localized basis (weak noise) or in the delocalized basis (strong noise), at a speed which is at least $\propto t^{-n}$.

\item[--] {\bf Novelty of idea and method. } The literature on the dynamics of open quantum systems and in particular on decoherence, is huge. The topic is treated by physicists, chemists and mathematicians (more recently it plays a role also in applications to biology \cite{moh}). To our knowledge, the dynamics of qubits generated by a random time independent Hamiltonian, as considered here, has not been analyzed before, even though it is a very natural model, akin to the famous Anderson model (see the introduction). The core technical idea underlying our work is that the irreversibility of the qubit dynamics can be quantified by the smoothness of the noise using the Riemann-Lebesgue Lemma. To the best of our knowledge, its connection to the dynamics of open quantum systems and in particular to the theory of decoherence, has not been exploited before. Combined with the explicit diagonalization of the qubit's time dependent $2\times 2$ density matrix we are able to carry out a direct analysis of the evolution. We are not aware that the expressions obtained in this work, for instance Lemmas \ref{lemma3} and \ref{analemma}, have been given or exploited before. We believe that our technically quite simple and direct approach might lead to further insights, also into the decoherence and entanglement of multi-qubit registers.

\item[--] {\bf Classical and quantum noises, markovianity.} There are some qualitative differences in the dynamical behaviour of a qubit subjected to different types of noise. Here are some characteristics described by our model:
\begin{itemize}
\item[C1] The speed of irreversible processes is polynomial in time, $\propto t^{-n}$ ({\em c.f.} \eqref{r1}).

\item[C2] This speed depends on the smoothness and the presence of infrared modes in the noise ({\em c.f.} \eqref{mainres'}).

\item[C3] Both the populations and coherences evolve in time (diagonal and off-diagonal density matrix elements in the energy basis), and the final state depends on the initial state ({\em c.f.} \eqref{finalstate}).
\end{itemize}

One of the trademarks of open system dynamics in the markovian approximation is the exponential time decay (approach of final state, decoherence). The associated characteristic rates are expressed in terms of the Fourier transform of the reservoir correlation function (the so-called spectral density of the noise). Clearly C1 deviates from this behaviour. A deviation of exponential time decay in non-markovian settings is known \cite{AL, Joos, BP, GZ, Haroche, PSE,Ferialdi}, but we have not yet analyzed the markovian properties of our current model. Both in quantum and classical stochastic (time-dependent) noise models, it is clear how to define the noise correlation function (which in turn determines decay rates). The analogue in our random matrix model is not obvious -- but it should be related to the smoothness of the probability density, as the latter determines the speed of the processes. Our property C2 indicates that the absence of slow noise modes accelerates the approach to the final state. Interestingly, however, the slow noise modes in quantum noise models have been shown to be necessary for {\em full decoherence} of the qubit \cite{MDimer}. In many models for open quantum system dynamics there is a unique stationary (final) state, typically an equilibrium state, and all initial states converge to it. In the current model this is not so, see C3. The emergence of a multitude of stationary states (and consequently a final state which depends on the initial one) has been reported in models with quantum noise due to two reasons: (i) the system-reservoir interaction is  energy-conserving, which leaves the populations of the system constant in time and only allows for decoherence \cite{PSE}, and (ii) the Hamiltonian has degenerate energy levels and shows some symmetry, which causes the existence of a manifold of stationary states for the coupled dynamics \cite{MBS-deg}.

\end{itemize}

\section{Proofs}
\label{proofsect}

The proofs are based on the explicit diagonalization of the Hamiltonian. Let $z\in\mathbb C$ and $a,b\in\mathbb R$. We diagonalize of the $2\times 2$ self-adjoint matrix to find its spectral representation,
\begin{equation}
H=\begin{pmatrix}
a & z\\
\bar z & b
\end{pmatrix} = \sum_{j=1,2}\lambda_j |\Psi_j\rangle \langle \Psi_j|.
\label{hamiltonian'}
\end{equation}
We consider the case where 
\begin{equation}
(a-b)^2+|z|^2\neq 0,
\label{m9.1}
\end{equation} 
which is equivalent to $\lambda_1\neq\lambda_2$ (if equality holds in \eqref{m9.1} then $H$ is a multiple of the identity --  this situation holds on a set of measure zero with respect to the noise probability measures and is not relevant for the dynamics). Then the eigenvalues and eigenvectors have the explicit expressions
\begin{equation}
\Psi_j = c^{(j)}_1 \Phi_1 + c^{(j)}_2\Phi_2,\qquad j=1,2,
\label{eqn15} 
\end{equation}
with 
\begin{equation}
\lambda_j = \frac{a+b + (-1)^{j+1} \sqrt{(a-b)^2+4|z|^2}}{2}
\end{equation}
and
\begin{equation}
c^{(j)}_1 = \frac{z}{\sqrt{|z|^2 +(a-\lambda_j)^2}}, \qquad 
c^{(j)}_2 =\frac{\lambda_j-a}{\sqrt{|z|^2 +(a-\lambda_j)^2}}.
\label{m9}
\end{equation}

\medskip

The functional calculus implies $e^{-it H} =  \sum_{j=1,2} e^{-i t \lambda_j} |\Psi_j\rangle \langle \Psi_j|$. Upon setting $a = E_1 +\xi_1$, $b = E_2 +\xi_2$ and $z=\xi_0$ in \eqref{hamiltonian'}-\eqref{m9}, a direct calculation yields the  expressions of the density matrix elements \eqref{m4}, as shown in Lemma \ref{lemma3} below. To express them, we  define the following functions of a (for now real) variable $P\neq 0$.
\begin{eqnarray}
Q(P) &=& \frac{1+ \sqrt{1+P^2}}{P}, \label{q6P}\\
h(P) &=&Q \frac{ (2\rho_{11}(0)-1)Q  + \rho_{21}(0) - \rho_{12}(0)Q^2}{(1+Q^2)^2},\nonumber\\
g_1(P) &=& - Q^2 \frac{(2\rho_{11}(0)-1)Q -\rho_{21}(0)- \rho_{12}(0)Q^2 }{(1+Q^2)^2},\nonumber\\
g_2(P) &=& \frac{(2\rho_{11}(0)-1)Q  - \rho_{21}(0)Q^2 + \rho_{12}(0)}{(1+Q^2)^2}.
\label{q2}
\end{eqnarray}
Here, $\rho_{ij}(0)$ are the matrix elements of the initial density matrix $\rho(0)$. We introduce real variables $x,y$ and set
\begin{equation}
P(x,y)=\frac{2x}{\varepsilon+y},
\label{q6}
\end{equation}
so that $P, Q$ and $h$, $g_1$, $g_2$ become functions of $x,y$. When taking averages (later on), we have in mind the convention \eqref{notation} for the meaning of $x,y$. 

\begin{lem}
\label{lemma3}
Consider a realization where the random variables are `frozen', {\em i.e.},  $\xi_{\rm o}=x$ and $\xi_{\rm d}=\xi_1-\xi_2=y$, for some fixed $x,y$ satisfying $x\neq 0$ or $y\neq -\varepsilon$ (so that \eqref{m9.1} holds).  Then the density matrix elements $\rho_{11}(t)$ and $\rho_{12}(t)$, defined in \eqref{m4}, are given by
	\begin{align}
	\label{q1}
	\rho_{11}(t) = &\  \frac{2Q^2}{(1+Q^2)^2}+   \Big( \frac{1-Q^2}{1+Q^2} \Big)^2 \rho_{11}(0) 
	-2  \frac{Q(1-Q^2)}{(1+Q^2)^2} {\rm Re} \rho_{12}(0)\nonumber\\ &+2{\rm Re}\ e^{-it(\varepsilon+y)\sqrt{1+P^2}} h(x,y) 
	\end{align}
	and
	\begin{align}
	\rho_{12}(t)= & \  \frac{Q(1-Q^2)}{(1+Q^2)^2} \big(1-2 \rho_{11}(0)\big)  
	+  \frac{4Q^2}{(1+Q^2)^2}\  {\rm Re} \rho_{12}(0)\nonumber\\  
	& + e^{-it(\varepsilon+y)\sqrt{1+P^2}} g_1(x,y) + e^{it(\varepsilon+y)\sqrt{1+P^2}} g_2(x,y).
	\label{q3}
	\end{align}	
\end{lem}
Since $\rho(t)$ is self-adjoint and has unit trace,  Lemma \ref{lemma3} specifies $\rho(t)$ entirely. We also remark that to arrive at the above formulas, we have used that $\varepsilon +y\ge 0$. 
\medskip

The time decay properties of the expectation of \eqref{q1} and \eqref{q3} are determined by the oscillating phases and the smoothness of the functions $h$ and $g_{1,2}$.
\begin{lem}  
\label{analemma}
The functions $h$, $g_1$ and $g_2$ have analytic extensions to $P \in {\mathbb C}\backslash\{\pm i [1,\infty)\}$. Their Taylor series at the origin (radius of convergence $1$), satisfy
\begin{eqnarray}
h(P) &=& -\frac{1}{2} \rho_{12}(0) P -\frac14 \big(1-2\rho_{11}(0)\big)P^2 +O(P^3),\label{e1}\\
g_1(P) &=& \rho_{12}(0)+\frac12\big(1-2\rho_{11}(0)\big) P +O(P^2),\label{e2}\\
g_2(P) &=& -\frac14 \rho_{21}(0) P^2 -\frac18 \big(1-2\rho_{11}(0)\big)P^3 +O(P^4).\label{e3}
\end{eqnarray}
It follows in particular from \eqref{e1}-\eqref{e3} that
\begin{equation}
g_1(P) = -4\, \frac{\overline{g_2(\bar P)}}{P^2} +O(P^2) = -2 \frac{h(P)}{P} +O(P^2).
\end{equation}
\end{lem}

{\em Proof of Lemma \ref{analemma}.}  The square root in \eqref{q6P} extends analytically to $P\in\mathbb C$ for $P$ such that  $1+P^2\not\in (-\infty,0]$ to avoid the branch cut.  According to \eqref{q6P}, $Q$ is meromorphic in this region for $P$, with a simple pole at the origin, 
\begin{equation}
Q = \frac2P + \frac P2 + O(P^3).
\label{q30P}
\end{equation} 
The relations \eqref{e1}-\eqref{e3} follow then in a simple way from \eqref{q2}.\hfill $\blacksquare$

\subsection{Proof of Theorem \ref{theorem1}}

\subsubsection{Decay of the time-dependent parts}

Our notational convention for the meaning of $x,y$ is \eqref{notation}, namely, $x$ is the off-diagonal and $y$ is the diagonal noise variable.

\begin{prop}
\label{pro1}
Suppose  the conditions of Theorem \ref{theorem1} hold. Denote by $F(x,y)$ any of $h(P(x,y))$, $g_1(P(x,y))$ or $g_2(P(x,y))$, where $P(x,y)=\frac{2x}{\varepsilon +y}$ (see \eqref{q6}). Then there is a constant $C_n$  {\em s.t.}  for all $t\ge0$,
\begin{eqnarray}
	{\mathbb E} \big[ e^{-it(\varepsilon+\xi_{\rm d})\sqrt{1+P(\xi_{\rm o}, \xi_{\rm d})^2}} F(\xi_{\rm o},\xi_{\rm d}) \big] &\le& \frac{C_n}{1+t^n}. \label{q5}
\end{eqnarray}
\end{prop}

{\em Remark.}  We do not make any assumptions on the smoothness or the size of the support of $\mu_{\rm o}(x)$.

\bigskip

\noindent
{\em Proof of Proposition \ref{pro1}.\ } 
We write $F(x,y) = F(P(x,y))$ and start by noticing that  $F(P(x,y))$ is an analytic function of two variables \cite{Saks} in the domain 
$$
{\mathcal D} = \big\{(x,y)\in{\mathbb C}^2\ :\   |{\rm Im} x| <\varepsilon -\eta_{\rm d},\   \ |y|<  \varepsilon\big\}.
$$
$F$ is analytic in a bigger domain but $\cal D$ suffices for our purposes as we only need it to contain ${\rm supp}\mu_{\rm o}\times {\rm supp}\mu_{\rm d}\in{\mathbb R}\times{\mathbb R}$. It is obvious that $P(x,y)=\frac{2x}{\varepsilon +y}$ is analytic in $\mathcal D$. Then according to Lemma \ref{analemma}, $F$ is analytic in $x,y\in\mathbb C$ satisfying $P(x,y)\not\in \pm i [1,\infty)$. As one readily verifies, his latter condition is satisfied in for $(x,y)\in{\mathcal D}$. Since $F$ is analytic on $\mathcal D$, so are all the derivatives $\partial_x^k\partial_y^\ell F(x,y)$. Moreover, $\partial_x^k\partial_y^\ell F(x,y)$ is bounded on any compact set inside ${\mathcal D}$, for arbitrary $k,\ell\in{\mathbb N}\cup\{0\}$.

The expectation value \eqref{q5} reads
\begin{equation}
\label{01}
	{\mathbb E} \big[ e^{-it(\varepsilon+\xi)\sqrt{1+P^2}} F(\xi_{\rm o},\xi_{\rm d}) \big] =\int_{{\mathbb R}^2} e^{-it(\varepsilon+y)q(x,y)} F(x,y) \mu_{\rm o}(x)\mu_{\rm d}(y) dx dy,
\end{equation}
where 
\begin{equation}
\label{qiu}
q(x, y)=\sqrt{1+\frac{4x^2}{(\varepsilon+y)^2}}.
\end{equation}
We show below that
\begin{equation}
\label{eqn6'}
\sup_{x\in{\rm supp}\mu_{\rm o}}\, \sup_{ t\ge 0}\bigg|t^n\int_{\mathbb{R}}e^{-it(\varepsilon+y)q(x,y)}  F(x,y) \mu_{\rm d}(y) dy \bigg| <\infty.
\end{equation}
Here, ${\rm supp}\mu$ denotes the support of $\mu$. 
The result \eqref{q5} then follow from \eqref{01} and \eqref{eqn6'}. To prove \eqref{eqn6'} we start by noting (this is a little calculation) that
\begin{equation}
\label{a2}
te^{-it(\varepsilon+y)q(x,y)}=iq(x,y)
\partial_ye^{-it(\varepsilon+y)q(x,y)}.  
\end{equation}
We integrate by parts $n$ times in \eqref{eqn6'}, using \eqref{a2} and the fact that $\mu_{\rm d}(y)$ is compactly supported (which makes the boundary terms vanish), to get
\begin{align}
\label{a3}
\bigg|t^n\int_{\mathbb{R}}&e^{-it(\varepsilon+y)q(x,y)} F(x,y) \mu_{\rm d}(y) dy \bigg|  \nonumber\\
&=\bigg|\int_{\mathbb{R}}e^{-it(\varepsilon+y)q(x,y)}
\big(\partial_y\circ q(y)\big)^n F(x,y)\mu_{\rm d}(y)dy\bigg| \nonumber\\
&\le \int_{\mathbb{R}}\big|\big(\partial_y\circ q(y)\big)^n F(x,y)\mu_{\rm d}(y)\big|dy,
\end{align}
where $(\partial_y \circ q(y))^n$ is viewed as the {\em operator} acting by  $n$ times applying $\partial_y \circ q(y)$. Namely, for a function $f$ of $y$, 
$$
(\partial_y \circ q(y))^n f(y) = \partial_y \Big( q(y)\partial_y \big( q(y) \cdots \partial_y (q(y)f(y))\cdots \big)\Big).
$$
Here, for notational simplicity, we consider $x$ fixed and simply write $q(y)$ instead of $q(x,y)$. 

We are going to show that 
\begin{equation}
\sup_{x\in{\rm supp} \mu_{\rm o}} \sup_{y\in{\rm supp} \mu_{\rm d}} \big|(\partial_y \circ q(y))^n F(x,y)\mu_{\rm d}(y)\big| <\infty,
\label{02}
\end{equation}
where ${\rm supp}\mu$ is the support of $\mu$.  
Then, since $\mu_{\rm d}$ has compact support, which constricts the integration domain of \eqref{a3} to a compact set, the result \eqref{eqn6'} follows from \eqref{02}. To show \eqref{02}, we expand the operator $(\partial_y\circ q(y))^n$ using the product law for derivatives, giving that  $\big(\partial_y \circ q(y)\big)^n F(x,y)\mu_{\rm d}(y)$ is a sum of $\tfrac12 (n+2)!$ terms, each one of the form 
\begin{equation}
\label{theq}
q^{(i_1)}(y)\cdots q^{(i_\ell)}(y) \mu_{\rm d}^{(j)}(y)\partial_y^k F(x,y),
\end{equation} 
where $(\cdot)^{(r)}$ denotes the $r$th derivative w.r.t. $y$, and where the indices satisfy $1\le \ell \le n$ and $0\le i_1,i_2,\ldots,i_\ell,j,k \le n$. So it is enough to show that $q^{(k)}$, $\partial_y^kF$ and $\mu_{\rm d}^{(k)}$, $0\le k\le n$, are all bounded. 

$\bullet$ As discussed above, $q(y)(\equiv q(x,y))$, given in \eqref{qiu}, is analytic in $\mathcal D$. Thus all $y$ derivatives are bounded, uniformly in any compact subset of $\mathcal D$. It follows that 
\begin{equation}
\label{-2}
\sup_{x\in{\rm supp} \mu_{\rm o}} \sup_{y\in{\rm supp} \mu_{\rm d}} \max_{0 \le k \le n }|q^{(k)}(y)|<\infty.
\end{equation}

$\bullet$ Again, the analyticity of $F$ in $\mathcal D$ and the ensuing boundedness of all its derivatives on any compact subset in $\mathcal D$ immediately gives 
\begin{equation}
	\label{2}
\sup_{x\in{\rm supp} \mu_{\rm o} }\sup_{y\in{\rm supp} \mu_{\rm d}} \max_{0 \le k \le n }|\partial^k_y F(x,y)|<\infty.
\end{equation}

$\bullet$ Finally, $\mu_{\rm d}(y)$ is $n$ times continuously differentiable with compact support and so the derivatives are  bounded, 
\begin{equation}
\label{3}
\sup_{y\in{\rm supp} \mu_{\rm d}} \max_{0\le k\le n} |\mu_{\rm d}^{(n)}(y)| <\infty.
\end{equation}

Keeping in mind that the left hand side of \eqref{02} is a sum of terms of the form \eqref{theq}, we see that the estimates \eqref{-2}, \eqref{2} and \eqref{3} show the bound \eqref{02}. We have thus shown \eqref{q5}. This completes the proof of Proposition \ref{pro1}. \hfill $\blacksquare$

\subsubsection{Final state}

According to Lemma \ref{lemma3} and Proposition \ref{pro1}, the final state is 
\begin{equation}
	\label{fs1}
	\lim_{t\rightarrow\infty} {\mathbb E}[ \rho_{11}(t)] =2{\mathbb E}\Big[ \frac{Q^2}{(1+Q^2)^2}\Big]  +  {\mathbb E}\Big[\Big( \frac{1-Q^2}{1+Q^2} \Big)^2 \Big]\rho_{11}(0) 
	-2{\mathbb E}\Big[\frac{Q(1-Q^2)}{(1+Q^2)^2}\Big]
{\rm Re} \rho_{12}(0)  
\end{equation}
and
\begin{equation}	
\label{fs3}
	\lim_{t\rightarrow\infty} {\mathbb E}[\rho_{12}(t)]= {\mathbb E}\Big[ \frac{Q(1-Q^2)}{(1+Q^2)^2}\Big] \big( 1- 2 \rho_{11}(0)\big)
	 + 4{\mathbb E}\Big[\frac{Q^2}{(1+Q^2)^2}\Big] {\rm Re} \rho_{12}(0).
\end{equation}	
Combining \eqref{fs1} and \eqref{fs3} with the definitions
\begin{equation}
	\label{52}
\alpha = 2{\mathbb E}\Big[ \frac{Q^2}{(1+Q^2)^2}\Big],\quad \beta =  {\mathbb E}\Big[\Big( \frac{1-Q^2}{1+Q^2} \Big)^2 \Big], \quad \gamma = {\mathbb E}\Big[\frac{Q(1-Q^2)}{(1+Q^2)^2}\Big]
\end{equation}
yields the result \eqref{finalstate}. Note that for fixed $y$,  $Q\propto x$ is an odd function of $x$ and hence so is $Q(1-Q^2)$. It follows from \eqref{52} that if $\mu_{\rm o}$ is even, then $\gamma=0$.

\subsection{Proof of Theorem \ref{theorem1'}}

Recall once again our notational convention for the meaning of $x,y$, as given in \eqref{notation} ($x$ is the off-diagonal, $y$ the diagonal noise variable). We need to analyze the decay in $t$ of  $\mathbb{E}\big[e^{-it\varepsilon\sqrt{1+P^2} }F(x)\big]$, where $F(x)$ is either of $h(x,0)$, $g_1(x,0)$ or $g_2(x,0)$ and $P=\frac{2x}{\varepsilon}$.  We start by noticing that 
\begin{equation}
	e^{-it\varepsilon\sqrt{1+P^2}}
	=\frac{i}{2t} \frac{\sqrt{1+P^2}}{P}  \partial_x\big(e^{-it\varepsilon\sqrt{1+P^2}}\big).
	\label{q16}
\end{equation}
Upon integrating $n$ times by parts and using that the boundary term vanish (as $\mu_{\rm o}$ has compact support), we get
\begin{eqnarray}
	\label{m30.1}
\lefteqn{
	\mathbb{E}\big[e^{-it\varepsilon\sqrt{1+P^2} }F(x)\big]
	= \int_{\mathbb{R}}  e^{-it\varepsilon\sqrt{1+P^2}}\mu_{\rm o}(x) F(x)dx}\nonumber\\ &=&\Big(-\frac{i}{2t}\Big)^n \int_{\mathbb{R}}e^{-it\varepsilon\sqrt{1+P^2}}\partial_x \tfrac{\sqrt{1+P^2}}{P}\Big(\cdots \partial_x \tfrac{\sqrt{1+P^2}}{P}  \mu_{\rm o}(x)F(x)\Big)dx,\quad 
\end{eqnarray}
where we apply $n$ times the operator $\partial_x \tfrac{\sqrt{1+P^2}}{P}$ to $\mu_{\rm o}(x) F(x)$. Once again, it is clear from \eqref{q6P}, \eqref{q2} that $P\mapsto F(2P/\varepsilon)$ is a $C^\infty$ function on $\mathbb R$. So all derivatives of $F(x)$ are bounded on the support of $\mu_{\rm o}$. By expanding the $n$ fold action of the $x$ derivatives inside the last  integral of \eqref{m30.1} (or, simply counting powers), we obtain the following result: If $\mu_{\rm o}(x)$ has a zero of order $k=1,2,\ldots$ at $x=0$, then the last integral in \eqref{m30.1} is finite (integrable at the origin) provided $k\ge 2n-1$.

{\em Proof of statement 1. \!in the theorem:}  According to \eqref{e1}-\eqref{e3}, $h(P)\sim P\propto x$ for $x\sim 0$ and moreover, $\rho_{11}(t)$ only depends on $F=h$, see \eqref{q1}. So the diagonal of ${\mathbb E}[\rho(t)]$ converges at speed $1/t$ even if $\mu_{\rm o}(0)\neq 0$. Furthermore, if $\rho_{12}(0)=0$, then again due to \eqref{e1}-\eqref{e3}, all of $F=h$, $g_1$, $g_2$ are $O(x)$ for $x\sim 0$ and then all matrix elements of ${\mathbb E}[\rho(t)]$ converge at speed $1/t$. In either of these cases, we get $	\mathbb{E}\big[e^{-it\varepsilon\sqrt{1+P^2} }F(x)\big] \le C/(1+t)$.

{\em Proof of statement 2. \!in the theorem:}  For $\mu_{\rm o}(x)\sim x^k$ at $x\sim 0$, the integral \eqref{m30.1} is finite provided $k\ge 2n-1$ and so $\mathbb{E}\big[e^{-it\varepsilon\sqrt{1+P^2} }F(x)\big] \le \frac{C}{1+t^n}$. \hfill $\blacksquare$
\bigskip

{\bf Acknowledgements. } This research was supported by an NSERC Discovery Grant from the Natural Sciences and Engineering Research Council of Canada. We are grateful to an anonymous referee for valuable comments.

\end{document}